\documentclass[twocolumn,preprintnumbers,prl]{revtex4}
\usepackage{graphicx}
\usepackage{hyperref}
\newcommand{\vev}[1]{ \langle {#1} \rangle }

\begin{document}

\preprint{UCB-PTH-02/26, LBNL-50854}

\title{Realistic Dirac Leptogenesis}

\author{Hitoshi Murayama and Aaron Pierce}
\affiliation{Department of Physics, University of California,
                Berkeley, CA 94720, USA}
\affiliation{Theoretical Physics Group, Lawrence Berkeley National Laboratory,
                Berkeley, CA 94720, USA}
\date{\today}

\begin{abstract}
  We present a model of leptogenesis that preserves lepton number.  The
  model maintains the important feature of more traditional leptogenesis 
  scenarios: the decaying particles that provide the CP violation necessary 
  for baryogenesis also provide the explanation for the smallness of the 
  neutrino Yukawa couplings.  This model clearly demonstrates that, contrary
  to conventional wisdom, neutrinos need not be Majorana in nature in 
  order to help explain the baryon asymmetry of the universe. 
\end{abstract}
\maketitle

\section{Introduction} 
One interface between particle physics and cosmology is the attempt to 
provide an explanation for the observed baryon asymmetry in the universe.  
Leptogenesis represents one of the most attractive possibilities for the 
generation of this asymmetry.  The recent discovery of neutrino 
masses has further increased the credibility of this 
scenario.  In its original incarnation 
\cite{Fukugita:1986hr}, leptogenesis relies upon the decay of right-handed 
Majorana neutrinos to create lepton number, which is 
subsequently transformed into baryon number  by the electroweak 
$B+L$ anomaly.  This traditional scenario relies in an essential way on 
the breaking of lepton number by the Majorana right-handed neutrinos.  
The attractive feature of this model is that the right-handed neutrinos 
responsible for the generation of the lepton asymmetry are also responsible 
for the smallness of the observed neutrino masses through the 
see-saw \cite{see-saw} mechanism.

Since the original model of (Majorana) leptogenesis, there have been 
two important observations.  First, the provoking observation has been 
made that it is not necessary to break lepton number 
to have a theory of leptogenesis, and that leptogenesis could be 
accomplished in a theory with Dirac neutrinos \cite{Dick:1999je}.  
We will review this 
idea in the next section.  A disadvantage of 
this idea, relative to the traditional models of leptogenesis, is 
that it possesses no relationship between the mechanisms responsible for the 
generation of the lepton asymmetry and the smallness of the neutrino masses.  
The second observation was that, in supersymmetric theories, it is possible 
to explain the smallness of the neutrino Yukawa couplings by relating their 
presence to supersymmetry breaking \cite{Arkani-Hamed:2000bq}.  Combining 
these two ideas allows us to once again relate the generation of the lepton 
asymmetry to the smallness of neutrino masses.  

This brings Dirac leptogenesis on to a footing equal to that of the 
traditional Majorana leptogenesis models.  The only ingredient that
this mechanism requires beyond the usual leptogenesis scenario is 
a $U(1)_{N}$ symmetry, which forbids the bare Yukawa couplings between 
the left and right-handed neutrinos.

\section{\label{sec:neutrinogenesis}Review of Leptogenesis with Dirac
Neutrinos} 

Reference \cite{Dick:1999je} noted that, even in a theory that conserves 
lepton number, a CP violating decay of a heavy particle can result 
in a non-zero lepton number for left-handed particles, and an 
equal and opposite non-zero lepton number for right-handed particles.  For 
most standard model species, Yukawa interactions between the left-handed and 
right-handed particles are sufficiently strong to cancel these two stores of 
lepton number rapidly.  However, the interactions of a 
right-handed Dirac neutrino are exceedingly weak, and equilibrium between 
left-handed lepton number and right-handed lepton number will not be reached 
until temperatures fall well below the weak scale.  By this time lepton 
number has already been converted to baryon number by sphalerons.    

To see how this scenario works, imagine that a negative lepton number is 
stored in the left-handed neutrinos, while a positive lepton number 
of equal magnitude is stored in the right-handed neutrinos.  
Sphalerons act only on left-handed particles, violating $B+L$ while 
conserving $B-L$.  This means part of the negative lepton number stored 
in left-handed neutrinos can be converted to a positive baryon number 
by the electroweak anomaly.  The (now smaller in magnitude) negative lepton 
number 
stored in the left-handed neutrinos ultimately equilibrates 
with the positive lepton number stored in the right-handed neutrinos
only after the temperature of Universe drops below electronvolts.  
The processes responsible for equilibrating the right and left-handed
neutrinos conserve both $B$ and $L$ separately.  The ultimate result is 
a universe with a total positive lepton number and a total positive baryon 
number.

\section{Small Yukawa Couplings}
The basic program in this letter is to generate small Dirac Yukawa 
couplings by integrating out a heavy field following the methods of 
\cite{Froggatt:1978nt}.  The smallness of the Yukawa couplings will be 
explained by the large ratio between the scale of supersymmetry breaking 
and the heavy masses.  The key point is that the same heavy fields can
be responsible for the generation of the CP asymmetry. 

The heavy fields to be integrated out are three pairs of vector-like 
leptons, $\phi$ and $\bar{\phi}$, one pair per
generation of Standard Model particles.  These fields 
transform as doublets under $SU(2)_{L}$ \footnote{There is no obstacle 
to promoting these doublets to full $SU(5)$ ${\bf 5}+{\bf 5}^*$
multiplets to preserve 
the success of gauge coupling unification.  The matter content is
that of ${\bf 27}$ in $E_6$ if supplemented by additional singlets.}.
The decay of these heavy leptons will also provide the necessary CP violation 
for leptogenesis.  In order to have CP violation in this sector, it is 
sufficient to have two generations. 

We work in the context of the Minimal Supersymmetric Standard Model (MSSM) 
augmented by three generations of right-handed neutrinos.  We forbid bare 
Yukawa couplings, $L N H_{u}$, through the use of a $U(1)_{N}$ 
symmetry \footnote{Here we imagine $U(1)_{N}$ as a global symmetry.  
  In case of worries concerning 
  violation of global symmetries by quantum gravity 
  effects ({\it e.g.}, \cite{Coleman:1988cy}), 
  we note that there is no obstacle to gauging this symmetry.  All that is 
  required is to introduce appropriate matter to keep the $U(1)_N$ symmetry 
  anomaly-free.},
under which the $N$ has charge $+1$, while all the fields of the MSSM are 
uncharged.  We also add a gauge singlet, $\chi$ that breaks $U(1)_{N}$ 
when it acquires a vacuum expectation value (vev).  The 
field content of the model, along with the charges under $SU(2)_{L}$, 
$U(1)_{L}$, $U(1)_{N}$, and $U(1)_{Y}$ is shown in Table~1.  $U(1)_{L}$ is the 
standard lepton number, which remains a symmetry in this model broken
only by the $SU(2)_L$ anomaly.
\begin{table}
\begin{center}
\begin{tabular}{ccccc}
  Field       & $U(1)_{L}$ & $U(1)_{N}$ & $SU(2)_{L}$ & $U(1)_{Y}$   \\ \hline 
  $N$         &    $-1$    &     $+1$   &   {\bf 1}   &     0        \\
  $L$         &    $+1$    &     $0$    &   {\bf 2}   & $-\frac{1}{2}$ \\
  $H_{u}$     &    $0$     &     $0$    &   {\bf 2}   & $\frac{1}{2}$  \\
  $\phi$      &    $+1$    &     $-1$   &   {\bf 2}   & $-\frac{1}{2}$ \\
  $\bar{\phi}$&    $-1$    &     $+1$   &   {\bf 2}   & $\frac{1}{2}$ \\
  $\chi$      &     $0$    &     $-1$   &   {\bf 1}   &    0
\end{tabular}
\caption{The field content and quantum numbers of the model.}
\end{center}
\end{table} 
With these charge assignments, the most general renormalizable 
superpotential is:
\begin{equation}
  \label{eqn:superpotential}
  {\mathcal W} \ni \lambda N \phi H_{u} + h L \bar{\phi} \chi + 
  M^{\phi} \phi \bar{\phi},
\end{equation}
where we have suppressed generation indices.  Upon integrating out the 
heavy vector lepton pair, we get the following superpotential:
\begin{equation}
  {\mathcal W}_{\it eff} \ni \lambda h \frac{N H_{u} L \chi}{M^{\phi}}. 
\end{equation}
Next, we arrange for the $\chi$ field to take on a weak-scale vev.  
We can accomplish this, for example, 
through the use of an O'Raifeartaigh model of 
the type used for neutrino masses in \cite{Borzumati:2000mc}. 
This approach gives $\vev{F_{\chi}} \simeq m_{3/2} M_{\it Planck} \neq 0$ 
and $\vev{\chi}=0$ in the limit of global 
supersymmetry, but $\vev{\chi} \simeq 16\pi^2 m_{3/2}/\kappa^3 \neq 0$, 
where $\kappa$ is a dimensionless coupling constant, after supergravity 
effects are taken into account.  Because of the large $\vev{F_\chi}$,
left-handed and right-handed sneutrinos equilibrate quickly above the
weak scale.  However, the asymmetry stored in the right-handed
neutrino (fermion) remains intact.
Interesting collider phenomenology could result from the large 
$F_{\chi}$ \cite{Arkani-Hamed:2000bq}.  In any case, it is clear that the 
Dirac neutrino Yukawa couplings, $y_{\nu}$, will be suppressed by the 
ratio of the weak scale to the heavy masses:
\begin{equation}
  y_{\nu} \sim h \lambda \frac{\vev{\chi}}{M^{\phi}}.
\end{equation}
Because $\vev{\chi}$ does not have to be exactly at the electroweak
scale, it gives an additional freedom beyond the traditional Majorana
leptogenesis.  We note that a very similar superpotential was considered in 
\cite{Kitano:2002px}, with the vev of the $\chi$ field replaced with a 
hard mass.

\section{Lepton Asymmetry}
It remains to check whether this scenario can generate a sufficient baryon
asymmetry.  CP violation will enter the theory through the decay of
the $\phi$ and $\bar{\phi}$ particles.  There are equal contributions from 
the decay of the scalar and fermionic components.  For simplicity, we will 
concentrate on the decay of the scalars.  The leading order contribution 
to the CP violation in $\phi$ decay comes from the interference between the 
tree-level diagrams and the absorbative part of the one-loop wave function 
renormalization diagrams \cite{Kuzmin:1985mm,Liu:1993ds,Flanz:1996fb}.  In the
case where there are two ``generations'' of $\phi-\bar{\phi}$ pairs, it is 
possible to rotate away all but one physical phase.  We will consider this 
case in the following.  Additional generations of the $\phi-\bar{\phi}$ pairs 
will just allow for the possibility of additional baryon number 
generation.  In addition, the two generation case is a good 
approximation if the masses of the $\phi$ particles are reasonably 
well-separated.  We take the mass matrix, $M^{\phi}$, to be diagonal 
with elements $M_{1}$ and $M_{2}$.
The diagrams relevant to the calculation of the CP asymmetry are shown in 
in Figure 1.  Restoring the generation indices to 
Eqn.~(\ref{eqn:superpotential}), we have:
\begin{equation}
  \label{eqn:superpotential2}
  {\mathcal W} \ni \lambda_{i \alpha} N_{\alpha} \phi_{i} H_{u} + 
        h_{\beta i} L_{\beta} \bar{\phi}_{i} \chi + 
        M^{\phi}_{a} \phi_{a} \bar{\phi}_{a},
\end{equation}

\begin{figure}
\begin{center}
\includegraphics[width=\columnwidth]{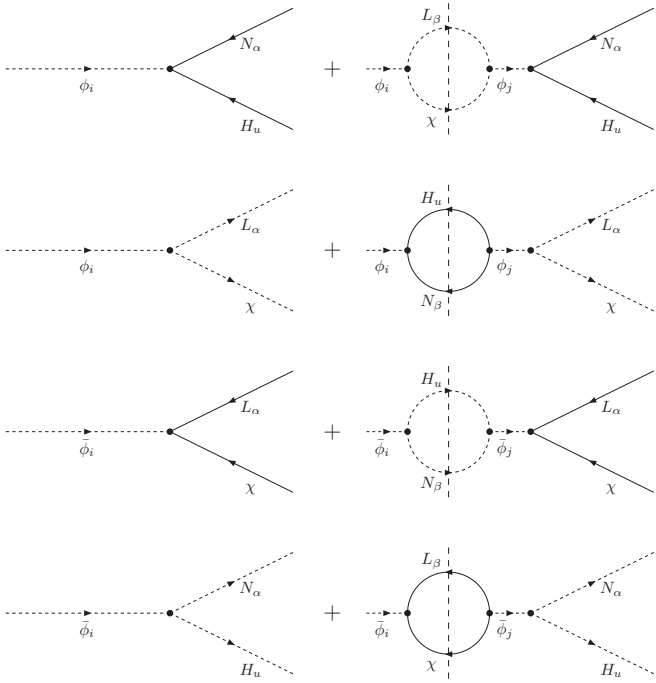}
\end{center}
\caption{Diagrams giving the leading contribution to the CP asymmetry 
        in $\phi$ and $\bar{\phi}$ scalar decays.  The absorbative part 
        of the one-loop diagrams contributes to the CP asymmetry.}
\end{figure}

Now we proceed with the calculation of the asymmetry.  In the case where 
the magnitudes of the masses $|M_{1}|$ and $|M_{2}|$ are well 
separated, the asymmetry will be dominated by the decay of the lightest 
$\phi-\bar{\phi}$ pair (we take $|M_{1}| < |M_{2}|$) and can readily 
be calculated (following the methods of \cite{Flanz:1996fb}).  
We now define the quantities
$J \equiv Im(h_{\beta 1}^{\ast} h_{\beta 2} 
        \lambda_{1 \alpha}^{\ast} \lambda_{2 \alpha} 
        M_{1} M_{2}^{\ast})$ and 
$\Delta M^{2} \equiv |M_{1}|^{2} - |M_{2}|^{2}$. In $J$, the $\alpha$ and 
$\beta$ indices run over the generations of the $L$ and $N$ particles.  For
the decay asymmetries, we find:
\begin{eqnarray}
  \epsilon_{\bar{N}} &\equiv &
  \frac{\Gamma(\phi_{1} \rightarrow N^{c} H_{u}^{c})-
    \Gamma(\phi_{1}^{c} \rightarrow N H_{u})}
  {\Gamma(\phi_{1})}\nonumber \\
  &=&
  -\frac{J}{4 \pi \, \Delta M^{2} \, 
    (|\lambda_{1 \alpha}|^{2} + |h_{\beta 1}|^{2})} \equiv \varepsilon; \\
  \epsilon_{L} &\equiv& \frac{\Gamma(\phi_{1} \rightarrow L \chi)-
    \Gamma(\phi_{1}^{c} \rightarrow L^{c} \chi^{c})}
  {\Gamma(\phi_{1})}= -\varepsilon; \\
  \epsilon_{\bar{L}} &\equiv& \frac{\Gamma(\bar{\phi}_{1} \rightarrow 
    L^{c} \chi^{c})-
    \Gamma(\bar{\phi}_{1}^{c} \rightarrow L \chi)}
  {\Gamma(\bar{\phi_{1}})} = \varepsilon; \\
  \epsilon_{N} &\equiv& \frac{\Gamma(\bar{\phi}_{1} \rightarrow N H_{u})-
    \Gamma(\bar{\phi}_{1}^{c} \rightarrow N^{c} H_{u}^{c})}
  {\Gamma(\bar{\phi_{1}})}= -\varepsilon. 
\end{eqnarray}
Note that $\Gamma(\phi) = \Gamma(\bar{\phi})$ due to supersymmetry,
because chiral superfields $\phi$ and $\bar{\phi}$ form a massive
super-multiplet.  Here we have used the same names for fermion and scalar 
fields in the same multiplet, and the $\alpha$ and $\beta$ indices 
labeling the generation of the final state particles are summed over.  
The above asymmetries in the decay amplitude give rise to a store of 
lepton number in the left-handed and right-handed (s)neutrinos.  
In the limit that the particles decay well out-of equilibrium 
(the ``drift and decay'' limit), the asymmetry is given by \cite{Kolb:vq}:
\begin{eqnarray}
N \equiv \frac{n_{N}}{s} 
\sim \frac{(\epsilon_{N} -\epsilon_{\bar{N}}) n_{\gamma}}{g_{\ast} n_{\gamma}} 
\sim \frac{-2 \varepsilon}{g_{\ast}} \\
L \equiv \frac{n_{L}}{s} 
\sim \frac{(\epsilon_{L} -\epsilon_{\bar{L}}) n_{\gamma}}{g_{\ast} n_{\gamma}} 
\sim \frac{-2 \varepsilon}{g_{\ast}}.
\end{eqnarray}
However, this limit is not necessarily applicable, as the condition for 
out-of-equilibrium decay, $\Gamma(\phi_{1})/ 2 H(M_1) \lesssim 1$,
is only marginally satisfied.  Therefore, one should solve the full 
system of Boltzmann equations numerically, including $2 \rightarrow 2$ 
scattering, to accurately determine the 
lepton asymmetry.  However, for an existence proof 
that this mechanism will work, we will not need to resort to these 
numerics: we simply note that for the specific choices of 
$\lambda= h^{T}$ and $\vev{\chi}$ equal to the electroweak vev, our 
asymmetry (and neutrino mass matrices) will reduce to that of the standard 
supersymmetric leptogenesis scenario with Majorana neutrinos.  It has been 
shown (for recent reviews see \cite{Buchmuller:2002xm}), that the generation 
of a sufficient lepton asymmetry is possible in this case, with the mass of 
heavy neutrinos at the $10^{10}$ GeV scale.  Indeed, it is possible that 
more complicated textures for $\lambda$ and $h$ might lead to a more 
efficient generation of a lepton asymmetry while remaining consistent with
low-energy data on neutrino oscillations.

\section{Cosmological and Astrophysical Constraints}
Theories of supersymmetric leptogenesis have tension with the gravitino 
problem; the reheat temperature must be low enough to avoid cosmological 
difficulties associated with gravitino production.  A typical
constraint is $T_{RH} \lesssim 10^9$--$10^{10}$~GeV for 1--2~TeV gravitino
\cite{Kawasaki:2000qr}.  On the other hand, 
the reheat temperature must be high enough to produce the particles 
(in our case the $\phi$ and $\bar{\phi}$) that need to be heavy in order 
to decay out of equilibrium.  However, as we have shown above, our 
scenario can reproduce a baryon asymmetry equal to that of the 
traditional leptogenesis scenario, which has been shown to be compatible 
with gravitino constraints \cite{Buchmuller:2002xm}.  There are a host of
other ideas to help with this tension.  For example, theories of anomaly 
mediation \cite{anomalymed}, have gravitino masses that are heavier than
the usual case by a loop factor, of order 100 TeV.  Furthermore, there 
has been recent work suggesting that it may be possible to significantly 
increase the mass of the gravitino in theories with weak scale 
supersymmetry, thereby obviating the gravitino problem \cite{Luty:2002ff}.  

Yet another possibility involves using coherent oscillations of the
scalar fields carrying lepton number \cite{Murayama:1992ua,Hamaguchi:2001gw}.  
In our case the $\phi=\bar{\phi}$ flat direction could be used, for example, 
with the O'Raifeartaigh model discussed earlier with $\kappa \sim 1$, 
$\vev{\chi} \sim 10$~TeV.  We make the assumption that $N$ and $L$ remain 
pinned to the origin.  If we stick to the
simplifying ansatz $\lambda = h^T$, we can scale $M^\phi$
proportional to $\vev{\chi}$ so as to reproduce the observed neutrino 
masses with the same Yukawa couplings as the traditional case.  This means 
that the CP asymmetry remains the same as well.
Working within the model of \cite{Hamaguchi:2001gw} (replacing $N$ with 
the $\phi=\bar{\phi}$ flat direction), in order to have the CP
asymmetry large enough, we require $M_1^\phi \gtrsim 10^{8}$~GeV.  This can 
well be consistent with the gravitino mass of $\sim 1$~TeV.  In addition, 
the possibility $\lambda \neq h^T$ gives even more freedom.

It would be interesting to study the gravitino problem with both
$\vev{\chi}$ and $\vev{F_\chi}$ (and hence the gravitino mass) as free
parameters, such as in models of gauge-mediated supersymmetry
breaking.  Smaller $\vev{F_\chi}$ gives a lighter gravitino, and the
constraint on the reheat temperature is more severe \cite{Moroi:1993mb}.  
However, smaller $\vev{F_\chi}$ allows smaller $\vev{\chi}$ while preventing 
the appearance of a negative eigenvalue in the sneutrino mass-squared
matrix.  This, in turn, would allow for lighter $\phi$, which helps 
with the gravitino problem.  Therefore, we expect Dirac leptogenesis
to accommodate models with lower $\vev{F_\chi}$ more easily than 
traditional leptogenesis models.

There might be a worry that the right-handed neutrinos 
could potentially represent a dangerous number of additional light
species at the time of Big-Bang Nucleosynthesis (BBN).  The constraint
is $\Delta N_\nu \lesssim 0.3$ \cite{Olive:1999ij}.
However, by the time of BBN, the contribution of right-handed neutrinos is
suppressed by the entropy factor: $\Delta N_\nu = 3 (T_{\nu_R}/T_{\rm
bath})^4 = 3 [g_{\ast}(1\mbox{MeV})/g_{\ast}({\rm MSSM}+N)]^{4/3} = 0.02$
and is safe.

When the $U(1)_N$ symmetry is broken by the $\chi$ vev or $F_\chi$
vev, a Nambu-Goldstone 
boson will be produced.  Generally, stringent astrophysical constraints 
on such particles (e.g.~Majorons, familons) are derived from looking at 
supernovae.  The usual constraints assume couplings between the SM fields 
and the Nambu-Goldstone bosons.  In contrast, in this case the right 
handed neutrino is the only light field charged under the $U(1)_{N}$.  
Consequently, the couplings of the Nambu-Goldstone bosons to the matter 
in the supernova will be exceedingly weak.  Nambu-Goldstone boson production 
processes will be suppressed by factors of $m_{\nu}/T$ relative to the 
usual case.  Since even the usual case (see, for example, 
\cite{Feng:1997tn}), can be made acceptable, there is clearly no problem 
here.

\section{Conclusion}
We have presented a realistic model of supersymmetric leptogenesis using Dirac 
neutrinos.  The smallness of the neutrino Yukawa couplings is related to 
the presence of heavy fields whose decay provides the seed for the baryon
number of our universe. The only ingredient used in this scenario above and 
beyond the usual leptogenesis scenario is the imposition of a $U(1)_{N}$ 
symmetry.  It would be interesting to search for a fundamental origin for 
this symmetry.  Because of the simplicity of this model, we believe that 
leptogenesis with Dirac neutrinos should be placed on an equal footing with 
the usual Majorana leptogenesis scenarios.

This model clearly displays that neutrinos need not be
Majorana in order for them to play a major role in the generation of the 
baryon asymmetry.  In this scenario, leptogenesis will not give rise to any 
signal in neutrino-less double beta decay experiments.

\begin{acknowledgments}
HM thanks Manfred Lindner for reminding him of Ref.~\cite{Dick:1999je}.
This work was supported in part by DOE under Contract
DE-AC03-76SF00098, and in part by the NSF under grant PHY-00-98840.
\end{acknowledgments}
                                

\end{document}